\documentclass[10pt,conference]{IEEEtran}
\IEEEoverridecommandlockouts
\usepackage{cite}
\usepackage{amsmath,amssymb,amsfonts}
\usepackage{algorithmic}
\usepackage{graphicx}
\usepackage{textcomp}
\usepackage{xcolor}
\usepackage{braket}
\usepackage{pdfpages}
\usepackage{color}

\def\BibTeX{{\rm B\kern-.05em{\sc i\kern-.025em b}\kern-.08em
    T\kern-.1667em\lower.7ex\hbox{E}\kern-.125emX}}

\makeatletter
\def\@citex[#1]#2{\leavevmode
\let\@citea\@empty
\@cite{\@for\@citeb:=#2\do
{\@citea\def\@citea{,\penalty\@m\ }%
\edef\@citeb{\expandafter\@firstofone\@citeb\@empty}%
\if@filesw\immediate\write\@auxout{\string\citation{\@citeb}}\fi
\@ifundefined{b@\@citeb}{\hbox{\reset@font\bfseries ?}%
\G@refundefinedtrue
\@latex@warning
{Citation `\@citeb' on page \thepage \space undefined}}%
{\@cite@ofmt{\csname b@\@citeb\endcsname}}}}{#1}}
\makeatother

\begin{document}

%\title{Conference Paper Title*\\

\twocolumn
\title{Circuit Implementation of Discrete-Time Quantum Walks on Complex Networks\\
}

\author{\IEEEauthorblockN{1\textsuperscript{st} Rei Sato}
\IEEEauthorblockA{\textit{AI division, Quantum Computing Group} \\
\textit{KDDI Research, Inc.,}\\
Fujimino, Ohara 2--1--15, Saitama, 356-8502, Japan \\
0000-0002-7878-7304}
\and
\IEEEauthorblockN{2\textsuperscript{nd} Kazuhiro Saito}
\IEEEauthorblockA{\textit{AI division, Quantum Computing Group} \\
\textit{KDDI Research, Inc.,}\\
Fujimino, Ohara 2--1--15, Saitama, 356-8502, Japan \\
0000-0003-4723-3357}
}

\maketitle

\begin{abstract}
In this paper, we propose a circuit design for implementing quantum walks on complex networks.  Quantum walks are powerful tools for various graph-based applications such as spatial search, community detection, and node classification.  Although many quantum-walk-based graph algorithms have been extensively studied, specific quantum circuits for implementing these algorithms have not yet been provided.  To address this issue, we present a circuit design for implementing the discrete-time quantum walk on complex networks.  We investigate the functionality of our circuit using the small-sized Watts-and-Strogatz model as the complex network model, comparing it with theoretical calculations.  This work offers a new approach to constructing quantum circuits for implementing quantum walks on arbitrary complex networks.
\end{abstract}


\section{Introduction}
%量子ウォークははいパフォーマンスtoolとして期待されている.  グラフ問題, 金融, AI, 最適化などに有望.
%現実問題をマッピングする場合、多くは複雑ネットワークとして存在する. 
%複雑ネットワークは正方格子などのregular latticeと違い, 固定した方向が存在しない. 
%そのため, これらの具体的な回路実装はない。
%In this paper, we propose a circuit design of quantum spatial search in complex networks.  

Quantum walks, the quantum counterparts of classical random walks, leverage the principles of superposition to exhibit powerful behaviors, making them highly useful for various graph-based applications such as spatial search algorithms~\cite{10.5555/1070432.1070590}, community detection~\cite{mukai2020discrete}, and node classification~\cite{sato2024qwalkvec}.  For example, they can be used to predict user interests and detect user communities in social networks.

For graph-based algorithms, datasets of complex networks are conventionally used~\cite{mukai2020discrete, sato2024qwalkvec}.  Complex networks are systems of interconnected nodes and edges that display intricate and often unpredictable behaviors.  These networks are characterized by non-trivial topological features~\cite{watts1998collective}, which differ from those of simple regular lattices, such as square lattices.

Although many algorithms combining quantum walks and complex networks have been extensively studied, specific quantum circuits for implementing these quantum algorithms have not yet been sufficiently provided. Therefore, in this work, we provide quantum circuits for implementing quantum walks on complex networks using IBM quantum simulators~\cite{Qiskit}. 
 We obtain ideal results using the Watts-Strogatz (WS) model with $8$ nodes.

%\section{Related work}
%回路設計に関しては~~がある。連続系でも。
%ページ足りない場合、いらない

\section{preliminaries}

We define a discrete-time quantum walk on a undirected complex network.  The complex networks $G(V,E)$ consists of a set of $N$ nodes $V=\{1,2,..,N\}$ and a set of edges $E=\{e_{ij}\}$.  The quantum state of the quantum walk on the complex networks with step $t$ is defined as
\begin{equation}
    \ket{\psi(t)} = \sum_{i=1}^{N}\sum_{j=1}^{k_i}\psi_{ij}(t)\ket{i}\otimes\ket{i\rightarrow j}.
\end{equation}
$\ket{\psi(t)}$ is defined in the Hilbert space $\mathcal{H} \equiv \mathcal{H}_N\otimes \mathcal{H}_k$, where $\ket{i}\in \mathcal{H}_N$ is associated with the positional degree of freedom, and $\ket{i\rightarrow j} \in \mathcal{H}_k$ is associated with the internal degree of freedom.  $\psi_{ij}(t)$ is the probability amplitude of $\ket{i\rightarrow j}$.  The time evolution of the quantum state $\ket{\psi(t)}$ is determined by the coin operator $\hat{C}$ and the shift operator $\hat{S}$:
\begin{equation}
    \ket{\psi(t)} = [\hat{S}\hat{C}]^t\ket{\psi(0)}.
\end{equation}
The initial state $\ket{\psi(0)}$ is given by
\begin{equation}
    \ket{\psi(0)} = \frac{1}{\sqrt{N}}\sum_{i=1}^{N}\frac{1}{\sqrt{k_i}}\sum_{j=1}^{k_i}\ket{i}\otimes\ket{i\rightarrow j}.
    \label{eq:initial_state}
\end{equation}
The coin operator is node-dependent since each node has a different number of links, i.e., 
\begin{equation}
    \hat{C} = \sum_i^{N} \ket{i}\bra{i}\otimes\hat{C}_i,
    \label{eq:coin}
\end{equation}
where $\hat{C}_i$ is given by $\hat{C}_i = 2\ket{s_i}\bra{s_i} - \hat{I}_i$, with $\ket{s_i} = 1/\sqrt{k_i}\sum_{j=1}^{k_i}\ket{i\rightarrow j}$. 
The coin operator $\hat{C}_i$ is a $k_i \times k_i$ matrix.
%Note that the numbering of the neighboring nodes $\{j_1, j_2, \cdots, j_{k_i}\}$ is arbitrary.  
The shift operator changes the position of the quantum walker based on the movement information of the nearest nodes. We define the shift operator as 
\begin{equation}
    \hat{S}\ket{i}\ket{i\rightarrow j} = \ket{j}\ket{j \rightarrow i}.
    \label{eq:shift}
\end{equation}
%The choice of the shift operator of Eq.~(\ref{eq:flip-flop}) is effective for the social networks with varying node degrees.
The probability of node $i$ is given by
\begin{equation}
    P_i(t) = \sum_{j=1}^{k_i}|(\bra{i}\otimes\bra{i\rightarrow j})\ket{\psi(t)}|^2.
    \label{eq:quantum_probability}
\end{equation}

\section{Research Problem}
\begin{figure}[t]
    \centering
    \includegraphics[width=88mm]{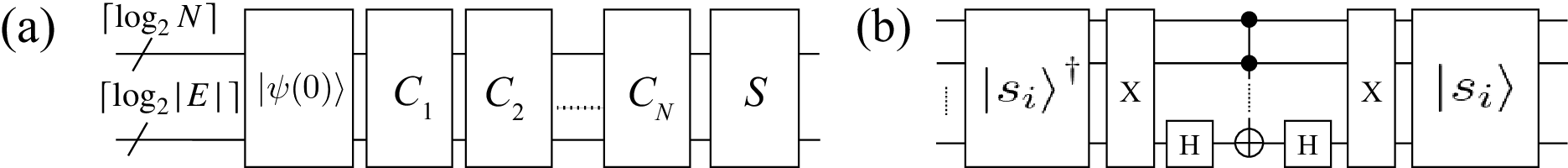}
    \caption{(a) Circuit implementation of discrete-time quantum walk on complex networks.  (b) The circuit design of coin operators $C_i$ of Eq.~(\ref{eq:coin}).  }
    \label{fig:proposed_circuit}
\end{figure}
%非正則グラフの量子回路の構築は困難です.  これは、各頂点のコイン空間が同じサイズではなくなることに加え, 量子を隣接ノードに移動させるためのシフト演算子の設計が複雑になるためです. これらの複雑なグラフがノードとサブノードの状態にどのようにラベルを付けるかが、最も効率的な量子回路実装を実現する上で重要な役割を果たす。この研究は、複雑なグラフ上で量子ウォークを実装するための量子回路の構築のためのアルゴリズムを提案する.  

%正方格子や超立方体, 完全グラフ, cycleグラフ上のDTQWではそれぞれのノードのコイン空間は同じサイズであるため比較的設計がしやすい. 一方で, 複雑ネットワークのような非正則グラフ上の場合, コイン空間の次元がノードごとに異なるため, 量子回路の設計が複雑になる.  また, シフト演算子もx,y,z方向のような規則的な法則をもたないため, エンコーディングが難しく, これまでに考えられていない. 

%本研究では, 初めて, 複雑ネットワーク上でDTQWが実装できる手法を提案する.  小規模の不規則なグラフをtoy problemとして, 理論通りに数値計算ができることを示す. 

The quantum circuit designs of the quantum walk on regular lattices are widely studied, such as one and two-dimensional lattice~\cite{acasiete2020implementation}, non-regular fractal~\cite{loke2012efficient}, hypercube~\cite{wing2023circuit}, complete graphs~\cite{wing2023circuit}, and cycle graphs~\cite{razzoli2024efficient}.  Despite these prior studies, quantum walk circuits on complex networks remain unresolved.  Constructing quantum circuits for complex networks is more challenging than for degree-regular graphs due to varying coin space sizes and different neighboring structures~\cite{loke2012efficient}.  It is important role to provide common framework of labeling the node and directions of quantum walkers on complex networks.

\section{proposed method}

%図1(a)は提案した量子回路の枠組みである.  この回路はノード情報を表現するための量子ビット数q_xと内部自由度を表現するための量子ビット数q_lからなる.  

%式(1)の初期状態準備では, 各量子状態はもつれ状態であり, 各ノード, エッジごとに確率振幅が異なるためゲート操作で埋め込むことが困難である.  したがって, QiskitのInitializaコマンドを使用して手動で回路に埋め込む. 

%式(4)(5)のコイン演算子とシフト演算子の実現のために, 量子ビット数q_lをエッジごとにラベル付けを行い, コントロール量子ビットとして使用する.  ただし, 式(4)のコイン演算子はノードごとにコイン演算子のサイズが異なるため, 固定された量子ビットq_lに対して異なるサイズのコイン演算子を作用させる必要がある.  その場合に, コイン演算子は図1(b)で示すように, 一般化Grover diffusionのを使用することで, 特定の内部自由度のみに回転操作を与える演算子を作成できるので, 式(4)を再現できる.   

%式(5)のシフト演算子では, エンコードされたエッジラベルをコントロールqubitsとして, q_xをターゲットqubitsとして使用することで, 式(5)のシフト演算子を再現することができる. 例えば, 図2(a)のノード0(=000)と7(111)の移動はエッジラベル000をコントロールqubitとして, 第1,2,3量子ビットをxゲートでフリップすることでノード間の移動を実現できる.  これは他のノードに対しても同様に動作する.
In this paper, we propose the design circuit for implementing quantum walk on arbitrary complex networks.  Figure~\ref{fig:proposed_circuit}(a) illustrates the framework of the proposed quantum circuit.  The circuit requires $q_x=\lceil \log_2{N}\rceil$ and $q_l=\lceil \log_2{|E|}\rceil$ qubits representing node position and their edge, respectively.

In the initial state preparation of Eq.~(\ref{eq:initial_state}), each quantum state is entangled, and the probability amplitude $\psi_{ij}$ differs for each node and edge, making it difficult to embed using gate operations.  Therefore, we manually embed the states into the circuit using Qiskit's `Initialize` command.

To implement the coin operators $C_i$ described in Eqs.~(\ref{eq:coin}), we have to apply different-sized coin operators to the fixed qubits $q_l$.  To address this issues, we propose using generalized Grover diffusion operators~\cite{generalized_grover} shown as Fig.~\ref{fig:proposed_circuit}(b) to perform rotational operations on specific internal degrees of freedom.

For the shift operator of Eq.~(\ref{eq:shift}), we use the encoded edge labels as controlled qubits and $q_x$ as target qubits.  For example, the movement between node $0$ (=$000$) and node $7$ (=$111$) in Fig.~\ref{fig:result_hist_DTQW}(a) is achieved by using the edge label $000$ as the controlled qubit and flipping the first, second, and third qubits with the $\rm X$ gate.  This method similarly applies to other nodes.

\section{Evaluation}
\begin{figure}[t]
    \centering
    \includegraphics[width=88mm]{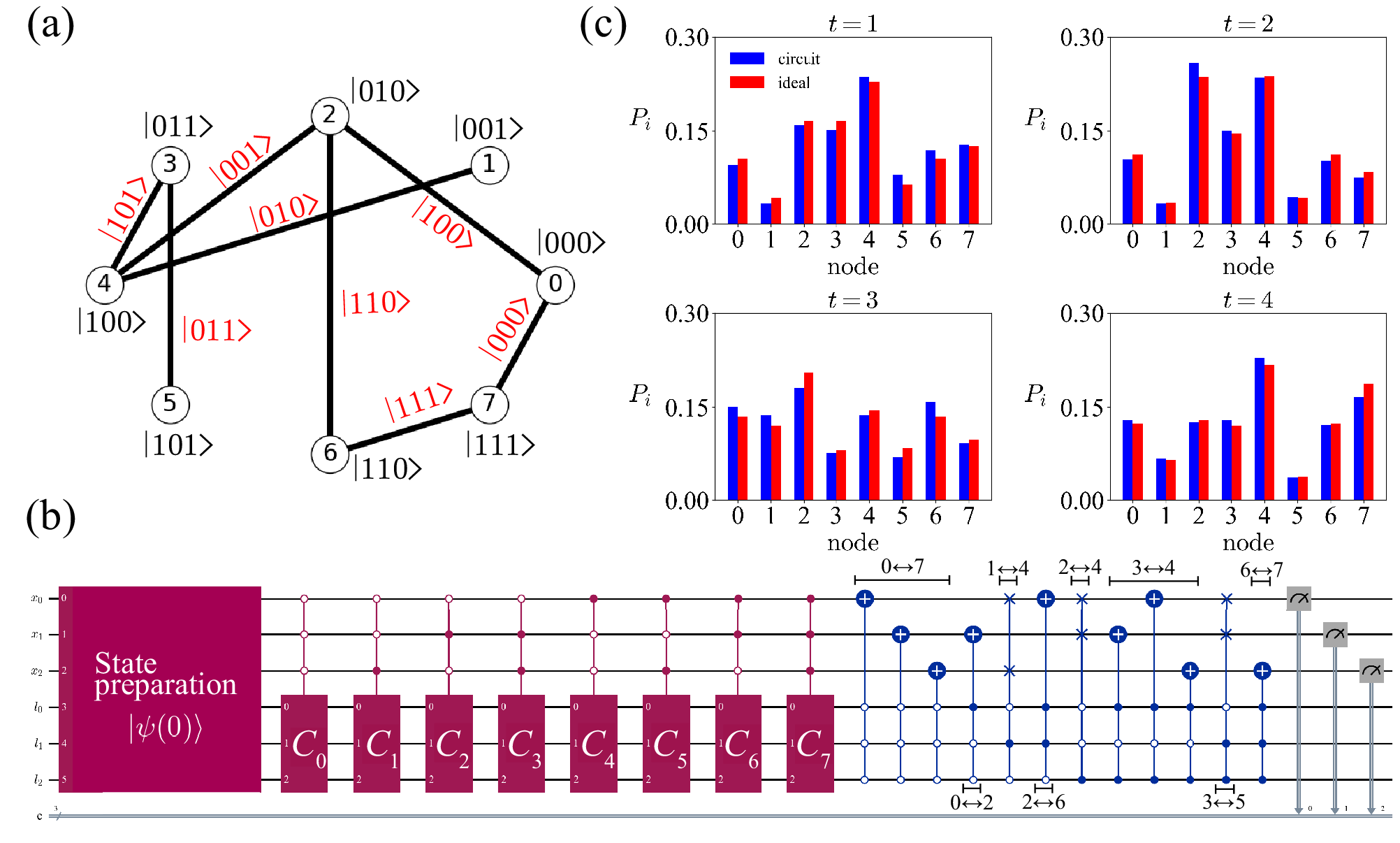}
    \caption{(a) WS model with $N=8, k=2, \beta=0.5$.  (b) The quantum circuit discrete-time quantum walk on WS model of Fig.~\ref{fig:result_hist_DTQW}(a) with $t=1$.  (c) The histogram of the probability of discrete-time quantum walk implemented by circuit simulation and theoretical calculation. }
    \label{fig:result_hist_DTQW}
\end{figure}
%量子探索を使用して回路設計の妥当性を評価する。
%\begin{table}[!t]
%\renewcommand{\arraystretch}{1.0}
%\caption{Results of the optimal iteration $Q$ and success probability $P$ for embedding all feasible solutions $\ket{T}$ of TSP.}
%\label{tab:result}
%\centering
%\begin{tabular}{|l|c|l|l|}\hline
%        & $N$ & Width &  Depth \\\hline\hline
%Our circuit & $4$ & $|N|+|E|$ & \\
%            & $8$ & $|N|+|E|$ & \\\hline
%\end{tabular}
%\end{table}
To validate our circuit, we use WS model~\cite{watts1998collective} as complex network data, where the model has parameters intensity of randomness $\beta$, node degree $k$, shown as Fig.~\ref{fig:result_hist_DTQW}(a).   Fig.~\ref{fig:result_hist_DTQW}(b) is an actual circuit for implementing the quantum walk on the WS model with $N=8$ and $t=1$.   The subscript $C_i$ is the coin operator of each node, and $ i \leftrightarrow j$ represents shift operators for node label $i$ and $j$.  Fig.~\ref{fig:result_hist_DTQW}(c) shows the probability of Eq.~(\ref{eq:quantum_probability}) for each step of the quantum walk on the WS model shown as Fig.~\ref{fig:result_hist_DTQW}(a).  We can see that our circuit works correctly by comparing theoretical simulation.

\section{Discussion}
Our proposed circuit requires $\lceil \log_2{N}\rceil+\lceil \log_2{|E|}\rceil $ width and more than $(\lceil \log_2{N}\rceil+\lceil \log_2{|E|}\rceil)t$ depth.  Our algorithm can also be applied to arbitrary real-world networks.  Therefore, it can be used for implementing quantum-walk-based graph processing algorithms on fault-tolerant quantum computers for tasks such as spatial search~\cite{10.5555/1070432.1070590}, community detection~\cite{mukai2020discrete} and node classification~\cite{sato2024qwalkvec}.

%Although our algorithm can be applied to regular lattices, the circuit depth may be larger compared to existing methods due to encoding all edges and using coin operators.  The efficient design of initial state encoding and shift operators remains a future work.

\section{Conclusion}
In this study, we proposed the quantum circuit for implementing the discrete-time quantum walk on complex networks.  We verified the functionality of our circuit using the WS model with $N=8$ as the complex network model.  This circuit can be used for quantum walk-based graph algorithms.

\section*{Acknowledgement}
A part of this work was performed for Council for Science, Technology and Innovation (CSTI), Cross-ministerial Strategic Innovation Promotion Program (SIP), “Promoting the application of advanced quantum technology platforms to social issues”(Funding agency : QST).

\bibliographystyle{unsrt}
\bibliography{IEEE}

\end{document}